\documentclass[twocolumn,showpacs,preprintnumbers,amsmath,amssymb,nofootinbib]{revtex4}

\usepackage{graphicx}
\usepackage{dcolumn}
\usepackage{bm}
\usepackage{graphics}
\usepackage{epsfig}
\usepackage{color}
\usepackage{amssymb}
\usepackage{amsmath}
\usepackage{amsfonts}

\begin{document}
\title{Nuclear Enthalpies}

\author{Jacek Ro\.zynek} \email{rozynek@fuw.edu.pl}
\affiliation{National Centre for Nuclear Research, Ho\.za 69, 00-681 Warsaw, Poland}

\begin{abstract}
We propose to benefit from a concept of the enthalpy in order to include volume
corrections to a nucleon rest energy, which are proportional to pressure and absent in a
standard Relativistic Mean Field (RMF) with point-like nucleons. As a result a nucleon
mass can decrease with  Nuclear Matter (NM) density, making an Equation of State (EoS)
softer. It is shown, how the EOS depends from nucleon sizes inside NM. The course of the
EoS in our RMF model agrees with a semi-empirical estimate and  is close to results
obtained from extensive DBHF calculations with a Bonn A potential, which produce the EoS
stiff enough to describe neutron star properties (mass--radius constraint), especially
the masses of ``PSR J1614–2230'' and ``PSR J0348+0432'', most massive ($\sim 2 M_\odot$)
known neutron stars. The presented model has proper saturation properties, including good
values of a compressibility.
\end{abstract}
 \maketitle
 Taking into account thermodynamic effects of
 pressure in finite volumes, we will describe how an energy per nucleon
 $\varepsilon_A=M_A/A$ and pressure evolves with NM density $\varrho$ in an RMF approach
 \cite{wa,ser,stocker,zm,GlenMosz}. The original Walecka version \cite{wa} of the linear RMF
in introduces two potentials: a negative scalar $g_SU_S$ and a positive vector
$U_V\!=\!g_V(U_V^0,\mbox{\textbf{\emph{0}}})$ fitted to a nuclear binding energy at the
equilibrium density $\varrho\!=\!\varrho_0$. The EoS for this linear, scalar-vector
($\sigma,\omega$) RMF model \cite{wa,ser} match a saturation point with too large
compressibility $K^{-1}\!=\!\varrho^2\frac{d^2}{d\varrho^2} \varepsilon_A\!\sim\!550MeV$
and is very stiff for higher densities, where the repulsive vector potential starts to
predominate the attractive scalar part. Nevertheless RMF models produce, after the
Foldy-Wouthuysen reduction, the good value of a spin-orbit strength at the saturation
density \cite{wa}. The dynamics of the potentials in the RMF approach are discussed e.g.
\cite{ms2} in four specific mean-field models \cite{wa,ser,stocker,zm}. In the ZM model
\cite{zm} a fermion wave function is re-scaled and interprets a new, density dependent
nucleon mass. It starts to decrease from $\varrho=0$ and at the saturation point
$\varrho\!=\!\varrho_0$ reaches 85\% of a nucleon mass $M_N$. But the nucleon mass
replaced at the saturation point by a smaller value would change the nucleon deeply
inelastic Parton Distribution Function (PDF) \cite{Jaffe}, shifting the Bj\"{o}rken $x
\propto (1/M_N)$. Such a shift means that nucleons will carry 15\% less of the
Longitudinal Momentum (LM), what should be compensated by the enhanced contribution from
a meson cloud for small $x<0.3$ to describe the EMC effect \cite{ms2,RW} in the RMF.
There is no evidence for a such huge enhancement \cite{newdata} in the EMC effect for
small x. Also the nuclear Drell-Yan experiments \cite{drell,ms2}, which measure the sea
quark enhancement, we described \cite{jacek} with a small 1\% admixture of nuclear pions
and the $M_N$ unchanged. Thus the deep inelastic phenomenology indicates that a change of
the nucleon mass at the saturation density is rather negligible. A nonlinear extension of
the RMF model \cite{stocker,hansel} assumes self-interaction of the $\sigma$-field with
the help of two additional parameters fitted to $K^{-1}\sim250MeV$ and an effective mass
$M_N^*=M_N+g_SU_S$. These modifications of a scalar potential give a softening of EOS
with a good value of compressibility. Modern RMF calculations \cite{Bielich,hansel} have
adjusted the EOS, fitting more mesons fields ($\rho$ for an isospin dependence) and
including the octet of baryons.

We propose to improve nuclear RMF models in a different way, namely by taking into
account volume contributions to a nucleon rest energy instead of a constant nucleon mass,
used so far in standard RMF models. Any extended object inside a compressed medium (like
a submerged submarine) needs an extra energy to preserve its volume. Thus from the
``deep" point of view, finite pressure correction should be taken into account in RMF
calculations with point-like nucleons, but also in the Quark-Meson Coupling (QMC) model
\cite{Hua}. To describe that dependence of a nucleon rest energy in a compressed medium
we will adopt a bag model. Considering  a role of finite nucleon sizes in compressed NM,
the simplest, original ($\sigma,\omega$) model \cite{wa,ser} with point-like nucleons,
which is too stiff, will be extended to get clear conclusions.

For fixed pressure and a zero temperature it is easy to show (see a first paragraph in a
next section), that definitions of a chemical potential $\mu$ or a Fermi energy, have the
same energy balance as an average, single particle enthalpy. An enthalpy contains in a
homogenous medium an interesting term, a work of a nuclear pressure $p_H$ in a
nuclear/nucleon volume, which will be investigated. It is the argument for our choice of
a Gibbs free energy with independent pressure $p_H$ in favor of a Helmholtz free one
(here an internal energy) with the volume, as an independent variable. Our results are
independent \cite{kumar} of that choice; like expressions on a chemical potential $\mu$
in (\ref{enthsing}).

We will neglect nuclear pion contributions above the saturation point. Dirac-Brueckner
calculations show that a pion effective cross section, in the reaction of two nucleons
$N+N=N+N+\pi$, is strongly reduced at higher nuclear densities above the threshold
\cite{hm1} ( also with RPA insertions to a self energy of $N$ and $\Delta$ \cite{oset}).
We restrict our degrees of freedom to interacting nucleons.
\section{Nuclear Enthalpy}
 \noindent At the beginning, let us consider effects generated by a volume of compressed NM.
Start with $A$ nucleons which occupy a volume $\Omega_A\!=\!A/\varrho$. They have to
perform a necessary work $W_A\!=\!p_H\Omega_A$ to keep a space $\Omega_A$ inside
compressed NM against nuclear pressure $p_H\!\doteq\!-(\partial M_A/
\partial \Omega_{A})$.
Thus interacting nucleons should provide not only the nuclear mass $M_A$, but rather
the nuclear enthalpy
\begin{eqnarray}
H_A\doteq M_A+W_A=M_A+A\ \!\frac{p_H}{\varrho}\label{enth}
\end{eqnarray}
which contains, besides the nuclear mass as an internal energy, the necessary work.
Taking appropriate thermodynamical derivatives with respect to $A$, we get following
relations between chemical potential $\mu$ and the enthalpy,
\begin{eqnarray}
\mu\doteq\!(\partial M_A/\partial A)_{_{\Omega_A}}\!\!\equiv\!(\partial H_A/\partial
A)_{{p_H}}\!=\!\varepsilon_{A}\!+\frac{\!p_H}{\varrho}\!=\!H_A/A \label{enthsing}
\end{eqnarray}
 for $A\rightarrow\infty$. Please note that  the same relation with pressure fulfills a nucleon Fermi energy
\begin{eqnarray}
E_F\!\doteq\!P_N^0(P_F\!)=\!(\partial M_A/\partial A)_{_{\Omega_A}}\!
 =\varepsilon_A\!+\!p_H/ \varrho=\!\mu\! \label{HvH}
\end{eqnarray}
of a nucleon with a Fermi momentum
 $P_F$; well-known as the  Hugenholtz-van Hove (HvH) relation \cite{kumar}, also proven in the self-consistent RMF approach
\cite{boguta}.

 The relativistic nuclear dynamics of nucleons in a nucleus, described by ``light cone"
momenta ($P^+_N,P^-_N,\textit{\textbf{P}}_N^{\bot}$), can be formulated
\cite{Jaffe,Fran,ms2} in the target rest frame, where $\textit{\textbf{P}}\!_A\!=\!0$.
In order to specify a total nuclear energy $P_A^0$  in compressed NM in a single
particle approach, let us discuss a longitudinal Momentum Sum Rules (MSR). Let's focus
our attention on the LM components $P^+_N\!=\!P_N^0\!+\!P_N^Z$ of $A$ nucleons. The
question is: do they add up to the internal energy $M_{A}$ or rather to the $H_A$,
greater then $M_A$ for positive pressure? To proceed our question let us look at a LM
distribution
\begin{eqnarray}
f_N(y)\!=\!\!\!\int\!\!{d^4P_N\over(2\pi)^4}\delta\left(y\!-\!{AP_N^+\over{P^+_A}}\right)
Tr\!\left[ {\gamma^+}G(P_N,P_A)\right],\label{structure}
\end{eqnarray}
with $y\!=\!AP_N^+/P_A^+$, which gives a Lorentz invariant fraction of a nucleon LM
$P^+_N$ in  the NM with a LM $P^+_A=P^0_A$. This distribution is manifestly covariant
and is expressed by a single nucleon Green's function \cite{Jaffe} $G(P_N,P_A)$ in the
nuclear medium, given e.q. in \cite{wa,ms2}. The trace is taken over the Dirac and
isospin indices and finally \cite{Mike,ms2}
\begin{eqnarray}
f_N\!\!\!\!&(\!\!\!&y)\!=\!\frac{4}{\varrho}\int\frac{S_N(P_N)d^4\!{{{P}}}\!_N}{(2\pi)^3}
\alpha\,\delta(y\!-\!AP_N^+/P^0_A) ; \label{RMF}
\end{eqnarray}
\noindent where a nucleon spectral function
\begin{eqnarray}
S_N=n(\mid\!\emph{\textbf{P}}_N\!\mid)\delta
(P_N^0-\sqrt{{M\!^{^*}_N}\!^2\!+\!{\emph{\textbf{P}}\!_N}\!^2} -g_V U_V^0) \nonumber
\end{eqnarray}
is given in the impulse approximation and $n$ is the Fermi distribution. Such a LM
distribution \cite{Jaffe},
 derived from matrix elements containing lower components
of a hadron wave function, includes a flux factor $\alpha\!=\!(1\!+\!{P_N^3}/{E^*_N})$
and thanks to this is properly normalized to the number of nucleons \cite{Fran}. After
integration (\ref{RMF}) the result is:
\begin{eqnarray} f(y)=\!\!&(\!\!\!&3/4)[P^0_A/(AP_F)]^3[
(AP_F/P^0_A)^2\!-\!(y\!-AE_F/P^0_A)^2]. \nonumber
\end{eqnarray}
where $y$ takes the values determined by the inequality
$(E_F\!-\!P_F)/P^0_A<(y/A)<(E_F\!+\!P_F)/P^0_A$. Integrating the LM fraction $y$ in NM

\begin{equation}
\!\int\!dyyf_N(y)\!=\!\frac{AE_F}{P^0_A}\!=\!A\frac{\varepsilon_A\!+\!{p_H}/{\varrho}}{P^0_A}=1,
\label{RMF3}
\end{equation}
and using HvH relation (\ref{HvH}) in a middle step we get the longitudinal MSR
(\ref{RMF3}) which gives a fraction of the nuclear LM taken by all nucleons
 \cite{Fran,ms2}; therefore equal 1.

Let us check it with the usual ``on mass shell" choice:
$P_A^0\!=\!M_A\!=\!A\varepsilon_A$. Then the MSR (\ref{RMF3}) is satisfied only at the
saturation point where $p_H=0$ \cite{ms2}. However, in the beginning we advocate to
choose the enthalpy $P_A^0\!=\!H_A\!=\!A\varepsilon_A\!+\!p_H\Omega_A$ as a total
nuclear energy. Taking (\ref{enthsing}) $H_A=A\mu$ we get
\begin{equation}
\!\int\!dyyf_N(y)=\!\frac{AE_F}{P^0_A}\!=\!\frac{AE_F}{H_A}\!=\frac{E_F}{\mu}=1 .
\nonumber \label{RMF2}
\end{equation}
Now the MSR (\ref{RMF3}) is always satisfied (\ref{HvH}) thanks to the finite volume
contribution $p_H\Omega_A$ to the nuclear energy. Thus we will use enthalpies, as
compact forms for total rest energies of nuclear or nucleon (parton) system.
\section{Nucleon Enthalpy}
\noindent We will discuss in a bag model, whether the nucleon mass $M_N$ or rather a
nucleon enthalpy $H_N$ should be, eventually, constant - independent from the density
inside the compressed medium. Such a question is absent in the standard RMF, where
nucleons are point-like with the constant mass $M_N$ independent of pressure inside NM.
But nucleons themselves are extended. In a compressed nucleon, partons (quarks and
gluons) have to do a work $W_N=p_H\Omega_N$ to keep a space $\Omega_N$ for a nucleon
"bag". It will involve functional corrections to a nucleon rest energy, dependent from
external pressure with a physical parameter - a nucleon radius $R$. Others modifications
connected with finite volume of nucleons, like correlations of their volumes, will be
neglected. The situation is similar to nucleons inside NM described in the previous
section, where we found that the MSR (\ref{RMF2}) is satisfied by the total energy
$P_A^0$ equal to the nuclear enthalpy $H_A$. Analogously, we introduce a nucleon enthalpy
$H_N$ with the nucleon mass $M_{pr}$ modified in the compressed medium
\begin{eqnarray}
H_N(\varrho)\doteq M_{pr}(\varrho)+p_H\Omega_N \ \ with \ \ H_N(\varrho_0)=M_N, \ \
\label{enthnuc}
\end{eqnarray}
 as a ``useful" expression for the total rest energy of a nucleon ``bag". Please
 note, that ``external" pressure $p_H$ used in (\ref{enthnuc}) is, of course,
 identical with nuclear pressure appearing in (\ref{enth},\ref{enthsing}).
Our volume corrections will change a nucleon rest energy but also will diminish
effectively a free space between nucleons for the given nuclear density, what modifies an
available space $\Omega_{A-}\!=\!(\Omega_A\!-\!A\Omega_N)$ and so nuclear pressure. Now
$p_H\!\doteq\!-{(\partial M_A/
\partial \Omega_{A-})}_A$.
A total enthalpy $H^T_A\!=\!H_{A-}+A(H_N\!-\!M_N\!)$ and using
(\ref{enth},\ref{enthsing},\ref{enthnuc}) we arrive to the HvH relation with extended
nucleons.
\begin{eqnarray}
H^T_A /\!A=\!\varepsilon_A\!-\!{(\partial M_A/
\partial \Omega_{A-})\!}_A/\!\varrho=\!\varepsilon_A\!+\!p_H\!/\!\varrho=\!E_F;  \label{enthtot}
\end{eqnarray}

\subsection{The nucleon mass in the Bag model in NM}
\noindent Describing nucleons as bags, pressure will influence their surfaces
\cite{Koch,Hua,bag,Kap,Jennings}. Finite pressure corrections to a mass can not be
described clearly by a perturbative QCD \cite{brown}. Let us discuss the relation
(\ref{enthnuc}) in the simple bag model where the nucleon in the lowest state of three
quarks is a sphere of a volume $\Omega_{N}$. Its energy $E_{Bag}$ is a function of the
radius $R_0$ with phenomenological constants - $\omega_0$, $Z_0$ \cite{Hua} and a
density dependent bag ``constant" $B(\varrho)$ with $B_0=B(\varrho_0)$. We have
\cite{MIT}
\begin{eqnarray}
E^0_{Bag}\!(R_0)\!\!&=&\frac{3\omega_0-Z_0}{R_0}+\frac{4\pi}{3}B(\varrho_0)R_0^3\propto~\!1/R_0
 , \label{bag}
\end{eqnarray}
The condition
\begin{eqnarray}
p_B=-\left(\partial E^0_{Bag}/\partial \Omega_{N}\right)_{surface} =0
\label{pressured2}
\end{eqnarray}
for pressure inside a bag in equilibrium,  measured on a surface, gives the relation
between $R_0$ and $B$, used in
 the end of (\ref{bag}). $E_{Bag}^{0}$ fits to the mass $M_N$ at equilibrium $p_H\!=\!p_B\!=\!0$.
 ($E^0_{Bag}$ differs from the $M_N$ by the c.m. correction \cite{Jennings}). In a
compressed medium, pressure generated by free quarks inside the bag \cite{MIT} is
balanced at the bag surface not only by intrinsic confining ``pressure" $B(\varrho)$
but also by nuclear pressure $p_H$; generated e.q. by elastic collisions with other
hadron \cite{Koch,Kap} bags, also derived in QMC model in a medium \cite{Hua}. In
equilibrium internal parton pressure $p_B$ (\ref{pressured2}) inside the bag is equal
(cf. \cite{Hua}), on a bag surface, nuclear pressure
\begin{eqnarray}
p_H\!=p_B\!\!&=&\!\!\frac{3\omega_0-Z_0}{4\pi
R^4}-\!B(\varrho)~~\rightarrow~~(B(\varrho)\!+\!p_H\!)R^4\!=\!const \nonumber
\label{pressure}
\end{eqnarray}
and we get the radius depending from $B\!+\!p_H$:
\begin{eqnarray}
R(\varrho)\!&=&\!\left[\frac{3\omega_0-Z_0}{4\pi (B(\varrho)+p_H(\varrho))}\right]^{1/4}.
 \label{rsolution}
\end{eqnarray}
Thus, the pressure $p_H(\varrho)$ between the hadrons acts on the bag surface similarly
to the bag ``constant" $B(\varrho)$. A mass $M_{pr}$ for finite $p_H(\varrho)$
 can be obtained from (\ref{bag},\ref{rsolution}):
\begin{eqnarray}
M_{pr}(\varrho)\!\!=\!\!\frac{4}{3}\pi\! R^3\!\left[4(B+p_H)\!-\!{p_H}\right]\!=
\!E_{Bag}^{0}\frac{R_0}{R}\!-\!p_H\Omega_{N}. \ \  \label{massbag}
\end{eqnarray}
 The scaling factor $R_0/R$ comes from
a well-known model dependence (\ref{bag}) ($E_{bag}^0\!\propto\!1/R_0$) in the spherical
bag \cite{MIT}. This simple radial dependence is now lost in (\ref{massbag}) and
responsible for that is the pressure dependent correction to the mass of a nucleon given
 by the product $p_H \Omega_N$. This term is identical with the work $W_N$ in
(\ref{enthnuc}) and disappear for the nucleon enthalpy
\begin{eqnarray}
H_N(\varrho)=E_{Bag}^{0}\frac{R_0}{R(\varrho)}\propto1/R(\varrho) . \label{enthbag}
\end{eqnarray}
\noindent  The nucleon radius $R(\varrho)$ reflects a scale of a confinement of partons.
Generally, for increasing $R(\varrho)$, $H_N(\varrho)$ (\ref{enthbag}) decreasing, thus
part of the nucleon rest energy is transferred from a confined region $\Omega_{N}$ to an
remaining space $\Omega_{A-}$ (\ref{enthtot}). For decreasing $R$, the $H_N$ increasing;
this allows the constant or increasing mass $M_{pr}$ (\ref{massbag}). Let us continue
with a ``conventional" nuclear case, when a nucleon interaction does not change an energy
of partons confined inside nucleons; therefore the enthalpy $H_N(\varrho)\!=\!M_N$ is
constant. Now, the constant $R$ (\ref{enthbag})
 require the work $W_N$ to keep the constant volume at the
expense of the nucleon mass $M_{pr}$ (\ref{massbag}). It is obtained (\ref{rsolution})
for the constant effective pressure
$B_{eff}\!\!=\!B(\varrho)\!+\!p_H(\varrho)\!=\!B(\varrho_0)$. The
$B(\varrho)\!=\!B(\varrho_0)\!-\!p_H$ gradually decreasing and disappears with
pressure in favor of strongly correlated colored quarks in the de-confinement phase
for $p_H\!\!=\!\!B(\varrho_0)\!\simeq60$ MeVfm$^{-3}$ \cite{MIT}, when
$\varrho\!\approx\!(0.5\!-\!0.6)$ fm$^{-3}$ (see FIG.1).

The internal pressure $B(\varrho)$, just as the external pressure $p_H(\varrho)$
(generated by an effective meson exchanges), has the same origin \cite{Bub} from an
interaction of quarks. Therefore, increasing $p_H(\varrho)$ we can expect the
corresponding decrease in $B(\varrho)$. Really, when pressure $p_H$ in NM is not taken
into account ($p_H=0$ in (\ref{rsolution})) the nucleon radius $R$, in the QMC model
\cite{Jennings}, increases in NM. However the nucleon radius $R$ is discussed in the
updated QMC model, which takes into account $p_H$ contributions \cite{Hua} to the bag
radius. They found this radius as a specific property of the  EoS, which depends from
the nuclear compressibility. In particular, for the ZM model \cite{zm}, which has the
realistic value of $K^{-1}\!\simeq\!225$ MeV, the nucleon radius remains almost
constant up to the density $\varrho\!=\!10\varrho_0$ (the volume corrections
(\ref{massbag}) to the nucleon mass are absent). However, for the stiff EOS of the
$(\sigma,\omega)$ model, they observe a strong increase of the nucleon radius up to
the density $\varrho=2\varrho_0$. Such an increase of the radius would diminish the
total rest energy $H_N$ (\ref{enthbag}) and the nucleon mass (\ref{massbag}), making
the EOS substantially softer - as a consistent feedback. Besides, in a Global Color
Symmetry Model (GCM) \cite{bag}, it has been shown that a decrease of the $B(\varrho)$
from the saturation density $\varrho$ up to $3\varrho$ by $~60$ MeVfm$^{-3}$ is
accompanied by a similar increase of pressure $p_H$.

Summarizing, the sum $B(\varrho)\!+\!p_H(\varrho)$ weakly depends on density in GCM or
QMC models with a reasonable stiff EOS, thus the bag radius remains about constant
(\ref{rsolution}). It justify our ``conventional" choice of the total nucleon rest energy
$H_N$, unchanged by an increasing NN repulsion. Just opposite to the case with the
constant nucleon mass $M_{pr}=M_N$, which requires the increasing total energy $H_N$
(\ref{enthnuc},\ref{enthbag}) and a decrease of the nucleon size. \vspace*{-0mm}
\section{Results and Discussion}
In the previous section we  argued for the constant total rest energy $H_N=M_N$, thus the
size of the nucleon is constant, regardless of pressure. We applied therefore following
formulas (\ref{enthnuc},\ref{enthtot}) for nucleon mass $M_{pr}$ inside NM:
\begin{eqnarray}
 M_{pr}(\varrho)\!&=&\!M_N-p_H(\varrho)\Omega_N, \ \ \ \ \ \varrho\geq\varrho_0 \label{enth1} \\
 p_H(\varrho)\!&=&\!\varrho^2 \varepsilon^{'}_A(\varrho)\!/(1\!-\!\varrho \Omega_N\!) \nonumber.
\end{eqnarray}
To carry out calculations we combine the $M_{pr}$ dependence (\ref{enth1}) of pressure
$p_H$ at the constant nucleon radius $R\!\!=\!\!R_0$, with the following standard
($\sigma-\omega$)
 RMF equations \cite{wa,ser} for the energy $\varepsilon_A$ in terms of the effective
mass $M_{pr}^*$:
\begin{eqnarray}
\varepsilon_A\!=\!\!&C&\!\!\!\!_1^2\varrho
\!+\!\frac{C_2^2}{\varrho}(M_{pr}\!-\!M_{pr}^*)^2\!\!+
\!\frac{\gamma}{\varrho}\!\!\int_0^{P_F}\!\!\!\frac{d^3\!\mbox{\emph{\textbf{P}}}\!_N}{(2\pi)^3}\sqrt{\mbox{\emph{\textbf{P}}}_N^2\!+\!{M_{pr}^{*2}}}
\nonumber\\
M^*_{pr}\!\!\!&=&\!\!M_{pr}\!-\!\frac{\gamma}{2C_2^2}\int_0^{P_F}\!\!\!\frac{d^3\!\mbox{\emph{\textbf{P}}}\!_N}{(2\pi)^3}
\frac{M^*_{pr}}{\sqrt{\mbox{\emph{\textbf{P}}}_N^2\!+\!M^{*2}_{pr}}}. \label{eq}
\end{eqnarray}
$\gamma$ denotes a level degeneracy and there are two (coupling) constants: a vector
$C_v^2$ and a scalar $C_s^2$,
 which were fitted \cite{wa,ser} at two different saturation points $(\varrho_0=0.16,0.19$ fm$^{-3}$ --
 see a figure caption) in NM. In a formula
$2C_1^2\!=\!C_v^2/M_N^2$, $2C_2^2\!=\!M_N^2/C_s^2$ with $g_VU^0_V\!=\!2C_1^2\varrho$,
$g_SU_S\!=\!M_{pr}-M^*_{pr}$ . \noindent
 Now the finite pressure corrections to the $M_{pr}$ (\ref{enth1})
 convert the recursive equations (\ref{eq}) above the saturation density $\varrho_0$
 to a differential-recursive set of equations, taking the general form
\begin{equation}
f(\varepsilon_A (\varrho),\varepsilon^{'}_A (\varrho))=0 ~for ~\varrho\geq\varrho_0.
\label{eq2}
\end{equation}

Note that (\ref{eq}) is obtained from the energy--momentum tensor for the model
Hamiltonian with a constant nucleon mass \cite{wa}. Here we assume that the same
equation with the mass $M_{pr}$ is satisfied in compressed NM. It should be a good
approximation, at least not very far from the saturation density.
\begin{figure}[t]
\vspace*{-25mm}
\hspace*{-6mm}
\includegraphics[height=12.9cm,width=9.8cm]{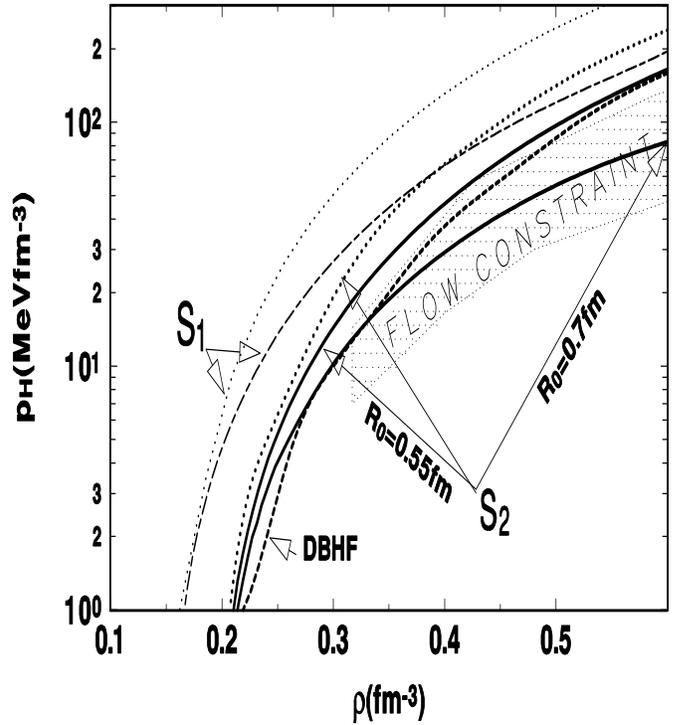}
\caption{Dotted lines show the pressure for NM, as a function of the density and the
constant nucleon mass $M_N$, for two different parameterizations of the ($\sigma,\omega$)
RMF model: $S_1$ \cite{ser} with $(\varrho_0=.16$ fm$^{-3}$) and $S_2$\cite{wa} with
$(\varrho_0=.19$ fm$^{-3}$). Long dashed line shows our results for constant nucleon
enthalpy $H_N$ with $S_1$ parameterization and a nucleon radius $R_0\!=\!0.7$ fm. Similar
results for set $S_2$, with $R_0\!=\!0.55$ fm or $R_0\!=\!0.7$ fm are marked as solid
lines. The area indicated by ``flow constraint'' taken from \protect\cite{pawel}
determines the allowed course of EoS, using an analysis which extracts from the matter
flow in heavy ion collisions from high pressure obtained there. The DBHF \cite{bonn}
calculation with a Bonn $A$ interaction is shown as a short dashed line.} \label{main}
\end{figure}

Linear ($\sigma\!-\!\omega$) models \cite{wa,ser}, with the constant mass $M_N$ produce
too stiff EoS; see FIG.1. Our results, which take into account nucleon volumes, are
compared with a semi-experimental estimate \cite{pawel} from heavy ion collisions and
indeed they correct the EOS, making it much softer. We have a good course of the EoS in
NM for the \{$R_0=0.7$ fm, set $S_2$\} up to the density $\varrho=0.6$ fm$^{-3}$. In
fact, below this density, a (partial) de-confinement is expected, which will change the
EoS above a phase transition \cite{quarkmatter}. For \{$R_0=0.55$ fm, set $S_2$\} the EoS
is relatively stiffer. However, it is a good candidate to investigate closely compact
stars \cite{last} in a case when hyperons will "soften" \cite{hansel,strange} the EoS
further. We see in the FIG.1 that both results for the set $S_2$ are rather close the
DBHF results, which produce the EoS able to describe \cite{NSTARS} the mass of ``PSR
J1614–2230'' or ``PSR J0348+0432''stars\cite{pulsar} (for $R_0=0.7$ fm slightly below the
DBHF for higher densities).  Alternatively, for an additional softening of the EOS
 the $S_1$ parametrization with our corrections (dashed
line) can be consider. It is worth mentioning that in a DBHF method there are
additional corrections \cite{bonn} from the self-energy, which diminish the nucleon
mass with density. In our model a volume part $p_H\Omega_N$ (\ref{enth1}) of the
constant total rest energy $H_N\!=\!M_N$ (\ref{enthnuc},\ref{enthbag}) effectively
diminishes the nuclear compressibility $K^{-1}$, changing its value from the
unrealistic $K^{-1}=560$ MeV \cite{wa} (set $S_2$) to the reasonable $K^{-1}=290$ MeV
obtained in our model for \{$R_0=0.55$ fm, set $S_2$\}. Other features of the Walecka
model, including a good value of the spin--orbit strength \cite{wa} remain unchanged
in our model.

The nucleon volume $\Omega_N$ is an important physical factor which strongly reduces
(\ref{enthtot}) the available space $\Omega_{A-}$. The relation (\ref{enthtot}),
$E_F\!=\!\varepsilon_A\!+\!p_H/\varrho$, connects the Fermi energy with nuclear
pressure $p_H$ acting in the  volume $\Omega_{A-}=(\Omega_A\!-\!A\Omega_N)$ and is met
with the $(0.1\!-3)\%$ numerical accuracy; worse for a higher density, ensuring
fulfillment of the MSR (\ref{RMF2}). This is a simple generalization of the HvH
relationship (\ref{HvH}) \cite{kumar} with finite-size nucleons.

\vspace{-0mm}
\section{Conclusions}
\vspace{-0mm} \noindent We have shown, how nucleon volumes in compressed NM affect
 the nuclear compressibility at equilibrium, reducing the nucleon mass and stiffness
of the EoS.
The compressibility \cite{pie} is lowered in linear ($\sigma\!-\!\omega$) model to the
acceptable value, giving the good course of EoS for higher densities. The nucleon mass
$M_{pr}(\varrho)$ (\ref{enth1}) occurred to be a pressure functional, what complements
the expression for a nuclear energy in our model. It effectively corresponds to
nonlinear, pressure dependent modifications of a scalar potential. Not accidentally,
in the widely used standard \cite{Bielich,hansel} RMF model with point-like nucleons
the good compressibility is fit by nonlinear modifications of a scalar mean field with
the help of two additional parameters. Thus, our results suggests to reconsider these
mean field parameters.

Particularly, when a nucleon ``confining" radius is constant in density, we have found
that the total rest energy $H_N$ of the nucleon is independent of density
(\ref{enthbag}), although the nucleon mass decreases with $\varrho$. Such a weak
dependence of $R$ from $\varrho$ is consistent with the phenomenological EOS.
The nuclear enthalpy (\ref{enth},\ref{enthtot}), as the total nuclear energy, satisfy
the longitudinal MSR (\ref{RMF2}) in the RMF approach. The presented model is suitable
for studying heavy ion collisions and neutron star properties (mass--radius
constraint); especially the most massive known neutron stars\cite{pulsar} recently
discover and we plan to include the octet of baryon, including strangeness, in a next
work.

This work is supported by a National Science Center of Poland, a grant
DEC-2013/09/B/ST2/02897.


\begin{thebibliography}{99}
\bibitem{wa} B. D. Serot and J. D. Walecka, Adv. Nucl. Phys. Vol. \textbf{16}
(Plenum, N. Y. 1986).
\bibitem{ser} R. J. Furnstahl and B. D. Serot, Phys. Rev. C
{\bf 41}, 262 (1990).
\bibitem{zm} J. Zimanyi and S.A. Moszkowski, Phys. Rev. C
{\bf 42}, 1416 (1990).

\bibitem{stocker}  J. Boguta and A.R. Bodmer, Nucl. Phys. A292, 413 (1977); J. Boguta, H. Stocker, Phys. Lett. B120, 289 (1983).
\bibitem{GlenMosz} N.K. Glendenning, S.A. Moszkowski,
Phys. Rev. Lett. \textbf{67}, 2414 (1991), N.K. Glendenning, F. Weber, S.A.
Moszkowski, Phys. Rev. C \textbf{45}, 844 (1992).

\bibitem{ms2} J.~R.~Smith and G.~A.~Miller, {Phys. Rev. C} {\bf 65}, 015211, 055206 (2002).

\bibitem{Jaffe} R. L. Jaffe, Los Alamos School on Nuclear
Physics, CTP 1261, Los Alamos, July 1985.
\bibitem{boguta} J.Boguta, Phys. Lett. 106B, 255, (1981).
\bibitem{hansel}  N.K. Glendenning, "Compact Stars",
Springer-Verlag, New York, 2000, P.Haensel, A.Y. Pothekin, D.G. Yakovlev, "Neutron
Stars 1", 2007 Springer.
\bibitem{RW} J. Ro\.zynek, G.Wilk, Phys. Rev. C \textbf{71},
068202 (2005).
\bibitem{newdata} J. Arrington, R. Ent, C.E. Keppel, J. Mammei, I. Niculescu, Phys.
Rev. C73, 035205 (2006), L.B. Weinstein et al., Phys.\! Rev.\! Lett.\! \textbf{106},
052301 (2011).
\bibitem{drell}D.M. Alde et al., Phys. Rev. Lett. \textbf{64}, 2479 (1990).
\bibitem{jacek} J. Ro\.zynek, Nucl. Phys. A \textbf{755}, 357c (2004).
\bibitem{Bielich} J. Schaffner-Bielich, M. Hanauske, H. St\"ocker, W. Greiner PRL, \textbf{89},
171101 (2002).
\bibitem{exvol} D.H. Rishke, M.I. Gorenstein, H. St\"ocker, W. Greiner,
Z. Phys. \textbf{51}, 485 (1991).
\bibitem{Hua} Guo Hua, J. Phys. G{\bf25}, 1701 (1999).
P.A. Guichon, Phys. Rev. Lett. B200, 235, (1988).
\bibitem{kumar} N.M. Hugenholtz and L.M. van Hove, Physica 24 (1958).
\bibitem{hm1} B. ter Haar and R. Malfliet, Phys. Rev. C {\bf 36}, 1611 (1987),
Phys. Rep. \textbf{149}, 287 (1987).
\bibitem{oset} E. Oset, L.L. Salcedo, Nucl. Phys. \textbf{468}, 631
(1987), "The Nuclear Methods and the Nuclear Equation of State", ed. M. Baldo, World
Scientific 1999.
\bibitem{Fran} L. L. Frankfurt and M. I. Strikman, Phys. Rep. 160, 235 (1988).
\bibitem{Mike} M. Birse, Phys. Lett. B, {\bf 299}, 188 (1993);
L. L. Frankfurt and M. I. Strikman, Nucl. Phys. B, 316(1989).
\bibitem{Koch} L. Ferroni and V. Koch, Phys. Rev. C \textbf{79}, 034905 (2009).
\bibitem{Jennings} X. Jin and B. K. Jennings, Phys. Rev. C \textbf{54}, 1427 (1996),
H. M\"{u}ller, and B. K. Jennings, Nucl. Phys. A \textbf{626}, 966 (1997).
\bibitem{Kap} J.I. Kapusta and Ch. Gale, "Finite
Teperatures Field Theory", Cambrdge Uniwersity Press, New York 2006.
\bibitem{bag} Y. Liu, D. Gao, H. Guo, Nucl. Phys. A695, 353 (2001);
R. T. Cahil, C. D. Roberts J. Praschifka, Ann. Phys. (NY), 188 (1988).
\bibitem{brown} G. E. Brown, M. Rho, Phys. Rev. Lett. \textbf{66}, 2720 (1991).
\bibitem{MIT} K. Johnson, Acta Phys. Pol. B6, 865 (1975),
A. Chodos et al., Phys. Rev. D \textbf{9}, 3471 (1974).
\bibitem{Bub} Buballa M., Nucl. Phys. A611,  393, (1996).
\bibitem{pawel} P.Danielewicz, R.
Lacey, W. G. Lynch, Science \textbf{298}, 1592 (2002).
\bibitem{bonn} T.
Gross-Boelting, C. Fuchs, A. Faessler, Nuclear Physics A \textbf{648}, 105 (1999); E.
N. E. van Dalen, C.Fuchs, A. Faessler, Phys. Rev. Lett. \textbf{95}, 022302 (2005);
Fuchs J. Phys. G \textbf{35}, 014049 (2008). (1995), D.P. Menezes et al., Phys. Rev. C
\textbf{76}, 064902, (2007).
\bibitem{quarkmatter} T. Kl\"ahn et al., Phys.Lett. B654, 170, (2007).
\bibitem{last} T. Kl\"ahn, D. Blaschke, R. Lastowiecki, Acta Phys. Pol. B Proc.
Suppl. 5, 757 - 772 (2012).
\bibitem{strange} I. Bombaci et al. Phys. Rep. 280 (1997).
\textbf{106}, 052301 (2011).\bibitem{NSTARS} T.~Kl\"ahn et al., Phys.\ Rev.\ C {\bf
74}, 035802 (2006).
\bibitem{pulsar} P. B. Demorest et al., Nature \textbf{467}, 7319 (2010),
Antoniadis et al., Science 340, 6131 (2013).
\bibitem{pie} J. Piekarewicz, Phys. Rev. C 64, 024307 (2001).

\end{thebibliography}
\end{document}